\documentclass[showpacs,twocolumn,floatfix,pra]{revtex4}
\usepackage{graphicx}
\usepackage{bm}
\newcommand{\ket}[1]{| #1 \rangle}

\newcommand{\rb}[1]{\left( #1 \right)}

\newcommand{\ew}[1]{\langle #1 \rangle}
\newcommand{\beq}{\begin{eqnarray}}
\newcommand{\eeq}{\end{eqnarray}}

\begin{document}
\title{A bipartite class of entanglement monotones for $N$--qubit pure states}
\author{Clive Emary}
\affiliation{Instituut-Lorentz, Universiteit Leiden, P.O. Box 9506, 2300 RA
Leiden, The Netherlands}
\date{\today}
\begin{abstract}
  We construct a class of algebraic invariants for $N$--qubit pure states based
  on bipartite decompositions of the system.
  We show that they are entanglement monotones, and that they differ
  from the well know linear entropies of the sub-systems.
  They therefore capture new information on the non-local properties of
  multipartite systems.
\end{abstract}
\maketitle

\section{Introduction}

In contrast to bipartite systems, the nature of entanglement
in multipartite systems is at present only partially understood.
An important step forward would be the determination
of all the algebraic invariants (AI) of a multipartite system.
For an $N$-particle pure state,
$\ket{\Psi} = \sum a_{b_1,\ldots ,b_N}\ket{b_1} \otimes \ldots \otimes \ket{b_N}$,
the AIs
are the set of algebraic functions of the coefficients $a_{b_1,\ldots ,b_N}$
which are invariant under local unitary transformations (LU) \cite{lin99}.

Although invariance under LU is crucial, it is not the whole story,
since one usually considers a more general situation in which the parties can
perform additional non-unitary operations (such as local measurements),
and can communicate classically with one another.  The combinations
of local operations and classical communication is denoted LOCC.
The pertinent measures of entanglement here
are the entanglement monotones (EMs).
These are AIs which are are non-increasing, on average, under LOCC \cite{vid00}.

Whether speaking of the AIs or the EMs,
the number required to entirely specify the non-local properties of the system
increases exponentially with the number of particles \cite{lin99}, and thus
a complete description seems out of reach for large systems.
Indeed, a complete set of AIs are only known for systems of up to 4 qubits \cite{bri03}.
Consequently, in considering multiparticle entanglement, we must seek
useful measures that capture essential features of the
entanglement, and/or are simple to calculate.

Emerging as the most important EM for pure states is the
hyperdeterminant $\Delta$
\cite{miy03}. The hyperdeterminant is afforded this status because it
is non-zero only for those states which possess genuine
$N$-particle entanglement.
For two qubits, $\Delta$ is the concurrence \cite{Woo98} and for three qubits,
the tangle \cite{Cof00}.  For systems of more than four qubits, the explicit
calculation of $\Delta$ is highly nontrivial.

Other useful EMs exist for $N$-qubit systems. A good
example is the von Neumann entropy (or its linearised form) of a single
qubit with the rest of the system.  This tells us whether the qubit in
question is separable or not.
Meyer and Wallach (MW) \cite{Mey02} introduced an $N$-qubit entanglement measure
which, as Brennen has shown \cite{bre03}, is equivalent to the
average of all the single qubit linear entropies.

MW construct these linear entropies in a particularly
elegant fashion, and in this article we introduce a new family of EMs
obtained from a generalisation of this construction.
We show that these quantities are EMs, and that they reflect an
aspect of the entanglement different to that captured by
the linear entropies of all sub-systems.
The utility of these EMs is demonstrated by considering the four qubit system.
Here, our EMs reproduce the fundamental algebraic invariants recently
described by Luque
{\it et al.} \cite{bri03}, and also prove useful in differentiating
between the nine families of four qubit entangled states recently described by
Verstraete {\it et al.} \cite{ver02}.
The simplicity of our construction gives the prospect of extending these
results to larger numbers of qubits.

The paper proceeds as follows.  In section
\ref{cons} we describe the construction of these entanglement monotones.
In section \ref{prop}, we consider some important properties; we demonstrate
that they are indeed EMs and compare them with linear entropies.
We consider in detail the four qubit system in section
\ref{four}, and conclude with a discussion in section \ref{disc}.
In the appendix we give the details of the proofs used here.

\section{Construction \label{cons}}

A pure state of $N$ qubits can be written as
\beq
  \ket{\Psi} = \sum_{b_1,\ldots , b_N} a_{b_1,\ldots ,b_N}
  \ket{b_1 \ldots , b_N}
  = \sum_{X=0}^{L-1} a_X \ket{X}
  \label{wfn1}
\eeq
where $X$ is the decimal for the binary string
${b_1,\ldots ,b_N}$ such that $0 \le X \le L-1$ with $L=2^N$.
Meyer and Wallach \cite{Mey02} introduced the single-qubit
``reduction operators'' $\iota^{(k)}_B$, which act on qubit $k$
in the following manner
\beq
  \iota^{(k)}_{B} \ket{b_1, \ldots , b_N}
  = \delta_{b_k,B}  \ket{b_1, \ldots ,\hat{b}_k, \ldots ,b_N}
  .
 \label{redop1}
\eeq
The circumflex denotes absence.  As $B \in \left\{0,1\right\}$ can take on one of two values,
the action of reduction operators at locus $k$ of $\ket{\Psi}$ gives the two vectors
\beq
  \iota^{(k)}_{0}\ket{\Psi}
  &=& \ket{\mathbf{V}_0^{(k)}}
  = \sum_{X=0}^{L/2-1}V_0^{(k)}(X)\ket{X}
  \\
  \iota^{(k)}_{1}\ket{\Psi}
  &=& \ket{\mathbf{V}_1^{(k)}}
  = \sum_{X=0}^{L/2-1}V_1^{(k)}(X)\ket{X}
\eeq
which MW combine to form the $N$ quantities
\beq
  D_1^{(k)} = 4 \sum_{X<Y} |V_0(X) V_1(Y) - V_0(Y) V_1(X)|^2
  .
  \label{d1}
\eeq
They then define their entanglement measure ${\cal Q}_1$ as the average
over all qubits $k$ of these quantities
\beq
  {\cal Q}_1
  &\equiv &
  \frac{1}{N} \sum_{k=1}^N D_1^{(k)}
  .
\eeq
As we see below, the quantities $D_1^{(k)}$ are themselves EMs, and are equal to the
linear entropies of the $k$th qubit \cite{bre03}.

We extend the MW construction by introducing reduction operators
$I^{(k_1, \ldots ,k_n)}_{B_{k_1}, \ldots , B_{k_n}}$ that act on $n$ qubits,
\beq
  I^{(k_1, \ldots ,k_n)}_{B_{k_1}, \ldots , B_{k_n}}
    \ket{b_1, \ldots , b_N}
  ~~~~~~~~~~~~~~~~~~~~~~~~~~~~~~~~~~~~~
  \nonumber\\
  \equiv
  \delta_{b_1,B_{k_1}},\ldots ,\delta_{b_n,B_{k_n}}
  \ket{b_1, \ldots ,\hat{b}_{k_1}, \ldots ,\hat{b}_{k_n} \ldots b_N}
  .
\eeq
We reference these operators by the locus
$\left\{k\right\} \equiv k_1, \ldots ,k_n$ describing
the qubits on which $I$ acts (the reduced qubits), and by
the decimal $X$ corresponding to the
bit-string $B_{k_1}, \ldots, B_{k_n}$.
The integer $n$ runs from unity to either $N/2$ or $(N-1)/2$
depending on whether $N$ is even or odd.
The action of $I^{\left\{k\right\}}_X$ on $\ket{\Psi}$ is to produce the
$(N-n)$ qubit state
\beq
  I^{(k_1, \ldots ,k_n)}_{X} \ket{\Psi}
  = \ket{\mathbf{V}^{(k_1,\ldots ,k_n)}_{X}}
  = \sum_{Y=0}^{\bar{L}-1} V^{(k_1,\ldots ,k_n)}_{X}(Y) \ket{Y}
\eeq
with $\bar{L} \equiv 2^{N-n}$.
For a given locus $ \left\{k\right\}$, there are $l \equiv 2^n$ vectors
$\mathbf{V}^{\left\{k\right\}}_{X}$ of length $\bar{L}$.

To construct our EMs we introduce the operator $dx_j$, which
assigns to vector $\mathbf{V}$ its $j^\mathrm{th}$ component,
i.e. $dx_j (\mathbf{V}) =V(j)$, and combine them in the
wedge product  defined as \cite{zwi96}
\beq
  \bigwedge_{i=0}^{l-1} dx_{j_i}  \rb{ \mathbf{V}_0, \ldots ,\mathbf{V}_{l-1}}
  \equiv
  \mathrm{Det} \rb{dx_{j_i}(\mathbf{V}_m)}_{i,m = 0,\ldots l-1}
  \label{WPdef}
  .
\eeq
The wedge product is completely
antisymmetric with respect to interchange of any two vectors in its argument,
and is zero for any two repeated arguments.

Writing the ordered set of vectors obtained from the action of all
$I^{\left\{k\right\}}_{X}$ operators at a given locus as
$\left\{\mathbf{V}\right\}
  \equiv \left\{ \mathbf{V}^{(k_1\ldots ,k_n)}_{0},\ldots,
    \mathbf{V}^{(k_1\ldots ,k_n)}_{l-1} \right\}$,
we define the quantities
\beq
  D^{(k_1,\ldots ,k_n)}_n
 \equiv
 l^2
\left\{
   \sum_{j_0<\ldots <j_{l-1}}
   \left|
     \bigwedge_{i=0}^{l-1} dx_{j_i}
     (\left\{\mathbf{V}\right\})
   \right|^2
\right\}^{2/l}
  .
\eeq
These are the objects that we study in the rest of the paper,
and as we show below, they are EMs.

For a given $n$ there are $N \choose n$ quantities $D^{(\left\{k\right\})}_n$, except when
$n=N/2$ with $N$ even, in which case there are half this number
since there are only $\frac{1}{2} {N \choose N/2}$ distinct bipartite
divisions.  For example, for four qubits we have four $D^{(k)}_1$ and three
$D^{(k_1,k_2)}_2$ measures.  These are not necessarily all independent.
For $n=1$, we recover the quantities of Eq. (\ref{d1}) introduced by MW.
Furthermore, for the two qubit system
$\ket{\Psi} = \sum_{ij=0}^1 A_{ij} \ket{ij}$,
we have $D_1 = 4|\mathrm{Det} A|^2 = {\cal C}^2$, the square of the
concurrence --- itself an EM.

\section{Properties \label{prop}}

\subsection{The quantities $D^{(\left\{k\right\})}_n$ are entanglement monotones}

In the appendix, we investigate the properties of the quantities
$D^{(\left\{k\right\})}_n$
under unitary transformations.  We show that not only are they
invariant under single qubit unitary transformations but,
writing the wave function as $ \ket{\Psi} = \sum_{i=0}^{l-1} \ket{\phi_i} \ket{V_i}$
where $\ket{\phi_i}$ are states of the $n$ reduced qubits, we show
that $D^{(\left\{k\right\})}_n$
is also invariant under unitary transformations of the whole subspace
spanned by $\ket{\phi_i}$.
Consequently, writing
$\ket{\Psi}$ in the Schmidt decomposition,
$\ket{\Psi}=\sum_{i=0}^{l-1} \ket{\phi_i}\ket{\mathbf{\tilde{V}}_i}$ with
$\ket{\mathbf{\tilde{V}}_i}$ orthogonal but not normalised,
leaves $D^{(\left\{k\right\})}_n$ unaltered.  
The Schmidt coefficients are
$\ew{\mathbf{\tilde{V}}_i|\mathbf{\tilde{V}}_i}  \ge 0$.

Using this decomposition,
we show in the appendix that $D^{(\left\{k\right\})}_n$ can be written as
\beq
  D^{(\left\{k\right\})}_n
  =
  l^2
  \left\{
    \prod_{i=0}^{l-1}
    \ew{\mathbf{\tilde{V}}_i|\mathbf{\tilde{V}}_i}
  \right\}^{2/l}
\label{dsd}
  .
\eeq
This form shows that $D^{(\left\{k\right\})}_n$ is indeed an EM.  From
Vidal \cite{vid00}, we know that all entanglement monotones of a bipartite system
can be expressed as $g\rb{\left\{ \alpha_i \right\}}$, where $g$ is a symmetric,
concave function of
the Schmidt coefficients $\left\{ \alpha_i \right\}$.
Equation (\ref{dsd}) shows $D^{(\left\{k\right\})}_n$ to be equal (up to normalisation) to the geometric mean
of the square of the Schmidt coefficients
( $g =\sqrt[l]{\alpha^2_1 \ldots \alpha^2_l}$).
Since for $n>1$ this is manifestly a concave function, it follows
immediately that $D^{(\left\{k\right\})}_{n>1}$ is an EM.  The quantities $D^{(k)}_{n=1}$
are also EMs, as can be seen by comparison with the linear entropy, below.

The power $2/l$ in Eq. (\ref{dsd}) is chosen to be the maximum that ensures that
$D^{(\left\{k\right\})}_n$ is an EM for all $n$.  This choice is justified further in the appendix, where
we show that with it, $D^{(\left\{k\right\})}_n$ transforms under local POVM in the same way as do
the concurrence-squared and the tangle.

These quantities have the interesting geometric interpretation as being
proportional to the square of the length of side of the hypercube with the same
volume as the parallelogram defined by the set of vectors
$\left\{ \mathbf{V}\right\}$.

\subsection{Comparison with linear entropies}
Since the construction of $D^{(\left\{k\right\})}_n$ is predicated on
a bipartite division of the system, we now compare
$D^{(\left\{k\right\})}_n$ with a more familiar EM based on the same
division, namely the linear entropy of qubits $\left\{k\right\}$ with the rest
of the system.  The linear entropies are defined as
\beq
  S^{(k_1,\ldots,k_n)}_n \equiv \eta_n
  \left[ 1- \mathrm{Tr}(\rho_{k_1,\ldots,k_n}^2)\right]
\eeq
where $\rho_{k_1,\ldots,k_n}$ is the reduced density matrix
of qubits $k_1,\ldots,k_n$, and
$\eta_n = 2^n/(2^n-1)$ provides suitable normalisation \cite{sco04}.

By utilising the Schmidt decomposition as above, we write
\beq
  S^{(\left\{k\right\})}_n = \eta_n
  \rb{
    1- \sum_{i=0}^{l-1}  \ew{\mathbf{\tilde{V}}_i|\mathbf{\tilde{V}}_i}^2
  }
\label{Snsc}
  .
\eeq
Thus, $S^{(\left\{k\right\})}_n$ is constructed
from the sum of the squares of the Schmidt coefficients,
and is an EM since $1-x^2$ is a concave function.

For $n=1$, this relation, plus the normalisation of
the Schmidt coefficients,
$\ew{\mathbf{V}_0|\mathbf{V}_0}+\ew{\mathbf{V}_1|\mathbf{V}_1}=1$,
shows that $D^{(k)}_1 = S_1^{(k)}$. This was the relation noted by Brennen
in connection with the original MW measure \cite{bre03}.
For $n \ge 2$ this identification does not hold.

\subsection{Simple applications}

Considering the form of $D^{(\left\{k\right\})}_n$ 
from Eq (\ref{dsd})., it is clear that
the minimum value of $D^{(\left\{k\right\})}_n$ is zero, occurring
when any $ \ew{\mathbf{\tilde{V}}_i|\mathbf{\tilde{V}}_i}=0$.  The
measure $D^{(\left\{k\right\})}_n$ is maximised when
$ \ew{\mathbf{\tilde{V}}_i|\mathbf{\tilde{V}}_i} =1/l$ $\forall i$,
from which we see that $D^{(\left\{k\right\})}_n$ is
normalised such that $\mathrm{max}~D^{(\left\{k\right\})}_n=1$.

If the block of qubits $\left\{k\right\}$ is separable from
the rest of the system, then
$D_n^{(\left\{ k \right\})}=S_n^{(\left\{ k \right\})}= 0$.
Moreover, if any of the qubits in $\left\{k\right\}$ separates then
$D_n^{(\left\{ k \right\})}=0$, whereas the corresponding entropy
is nonzero in general.  Neither $S_n^{(\left\{ k \right\})}$
nor $D_n^{(\left\{ k \right\})}$ are necessarily zero if there are separable
qubits
in the conjugate locus $\left\{\overline{k}\right\}$.

For the $N$-qubit GHZ (Greenberger-Horne-Zeilinger \cite{Gre89})  state,
$
  \ket{\gamma} \equiv 2^{-1/2}
  \rb{\ket{0}^{\otimes N} + \ket{1}^{\otimes N}}
$,
one has
$D^{(k)}_1=1$ and $D^{(\left\{k\right\})}_{n\ge 2}=0$.
For the $N$-particle $W$--state,
$
  \ket{\omega} \equiv N^{-1/2}
  \sum_{j=1}^N
  \ket{0}^{\otimes j-1} \otimes \ket{1} \otimes \ket{0}^{\otimes N-j}
$,
we find $D^{(k)}_1 =4(N-1)/N^2$ and $D^{(\left\{k\right\})}_{n\ge 2}=0$.
In contrast $S^{(\left\{k\right\})}_{n\ge2}$ is non-zero for both these states;
$S^{(\left\{k\right\})}_n(\gamma) = \eta_n/2$ and
$S^{(\left\{k\right\})}_n (\omega)= 2\eta_n (N-n)n /N^2$ for all $n$ \cite{sco04}.

A necessary, but not sufficient, requirement that at least one
$D^{(\left\{k\right\})}_n$ be nonzero is that $\ket{\Psi}$ must be a superposition of at least
$2^n$ states.
For example, the four qubit state
$1/2(\ket{0000} + \ket{0101} +\ket{1010}+\ket{1111})$ has
$ D_2^{(1,2)} = D_2^{(1,4)} = 1$, $ D_2^{(1,3)} = 0$.

Finally, to demonstrate that information is contained in $D^{(\left\{k\right\})}_{n\ge 2}$
that is not in $S^{(\left\{k\right\})}_{n\ge 2}$, consider the two 4--qubit states
\beq
  \ket{\psi_\pm} =
  \frac{1}{\sqrt{8}}
  \rb{
     \ket{0000} + \ket{0001} \pm \ket{0110} \pm \ket{0111}
  \right.
   \nonumber\\
  \left.
    \pm\ket{1001} + \ket{1010} + \ket{1100} + \ket{1111}
  }
 .
\eeq
Both states have the same linear entropies
but, whereas $\ket{\psi_+}$ has all $D^{(\left\{k\right\})}_2=0$,
$\ket{\psi_-}$ has $D_2^{(1,2)} = D_2^{(1,3)} =1/2$,
$D_2^{(1,4)} = 0$. This shows $D^{(\left\{k\right\})}_n$ to be independent of
$S^{(\left\{k\right\})}_n$ for $n\ge2$.

After these simple examples, we now consider in detail the case of four qubits.

\section{Four qubits \label{four}}

Consider the general four-qubit state
\beq
  \ket{\Psi} = \sum_{x=0}^{15} a_x \ket{x}
\eeq
in the usual decimal representation of a bit-string.
The explicit forms of $S^{(\left\{k\right\})}_1=D^{(\left\{k\right\})}_1$ and $S^{(\left\{k\right\})}_2$ are unenlightening.
However, the set of
three $D^{(\left\{k\right\})}_2$ operators reveals the use of this construction.
We find
\beq
  D_2^{(1,2)} &=& 16
  \left| \mathrm{Det}
  \rb{
  \begin{array}{cccc}
  a_0 & a_4 & a_8 & a_{12} \\
  a_1 & a_5 & a_9 & a_{13} \\
  a_2 & a_6 & a_{10} & a_{14} \\
  a_3 & a_7 & a_{11} & a_{15}
  \end{array}
  }
  \right|
\eeq
\beq
  D_2^{(1,3)}&=& 16
  \left| \mathrm{Det}
  \rb{
  \begin{array}{cccc}
  a_0 & a_2 & a_8 & a_{10} \\
  a_1 & a_3 & a_9 & a_{11} \\
  a_4 & a_6 & a_{12} & a_{14} \\
  a_5 & a_7 & a_{13} & a_{15}
  \end{array}
  }
  \right|
\eeq
\beq
  D_2^{(1,4)}&=& 16
  \left| \mathrm{Det}
  \rb{
  \begin{array}{cccc}
  a_0 & a_1 & a_8 & a_{9} \\
  a_2 & a_3 & a_{10} & a_{11} \\
  a_4 & a_5 & a_{12} & a_{13} \\
  a_6 & a_7 & a_{14} & a_{15}
  \end{array}
  }
  \right|.
\eeq
These are the moduli of the three fundamental algebraic invariants
found by Luque {\it et al.} for four qubits using
classical invariant theory \cite{bri03}.  The status of these algebraic invariants
is elevated to EMs once the modulus is taken.  Note that only two of these three
quantities are independent.

Furthermore, our measures $D^{(\left\{k\right\})}_n$ are useful in
distinguishing between types of entanglement in 4--qubit systems.
The states of
$N$ qubits may be grouped into families under the principle that
all the members of a family may be converted into one
another using LOCC with some finite, but not necessarily certain, probability
of success.  States connected in this way are said to be related by
SLOCC, standing for {\em stochastic} LOCC \cite{dur00, ver02,ver03}.
In deciding which family an arbitrary state belongs to,
we can consider the EMs. Since EMs are non-increasing under LOCC,
the property of a state having an EM equal to zero is preserved under
LOCC.  These zero EMs may therefore serve to differentiate between the
families.

Verstraete {\it et al.} \cite{ver02} have
analysed the properties of four qubit systems under SLOCC, and have
demonstrated that there are nine
distinct families of 4--qubit states.
The generic family of four qubits $G_{abcd}$ is identified as being the
only family with hyperdeterminant $\Delta \ne 0$ and this is the only family 
having
genuine 4--particle entanglement.
Of the remaining eight families, we can distinguish three different groups
based on the $D^{(\left\{k\right\})}_2$ measures.  The families $L_{abc_2}$ 
and $L_{ab_3}$ have no
zero $D^{(\left\{k\right\})}_2$.  Families $L_{a_2b_2}$ and $L_{a_4}$ have a 
single
zero $D^{(\left\{k\right\})}_2$, and the remaining four families have all
$D^{(\left\{k\right\})}_2=0$.  This
classification holds for all generic members of each family, but
may fail in special cases of zero measure,
such as the completely separable state that belongs to the generally
non--separable family $L_{abc_2}$.

It is clear that further EMs are required to complete this classification.
The linear entropies are not useful in this context, as they have no obvious
relation to the SLOCC families, although they may be used to identify separable
states.

In the context of SLOCC, we also mention the set of EMs introduced by 
Verstraete {\em et al.},
both for the four qubit system \cite{ver02}, and more generally \cite{ver03}.  
These
are different to the quantities introduced here, but share the interesting
similarity of being of the form of the modulus of the sum of products of the
wave function amplitudes $a_X$ (see Eq. (\ref{wfn1})) combined with 
antisymmetric tensors.  These
quantities bare a closer relation to the hyperdeterminant than do the measures
$D^{(\left\{k\right\})}_n$ \cite{ver03}.

\section{Discussion \label{disc}}

We have introduced the algebraic invariants $D^{(\left\{k\right\})}_n$, which are entanglement
monotones for pure states of $N$ qubits.  They arise from
considering all bipartitions of the system, and we have compared $D^{(\left\{k\right\})}_n$ with the
linear entropies $S^{(\left\{k\right\})}_n$ of the same partitions.

There are, in principle, an infinite number of EMs based upon
a given bipartite decomposition of an $N$-qubit state, as any concave, symmetric
function of the Schmidt coefficients is an EM.
The usefulness of the linear entropy as indicator of separability is clear,
and the $D^{(\left\{k\right\})}_n$ may be seen as complementary to the linear entropies.
Whereas $S^{(\left\{k\right\})}_n$ are based upon the sum of the squares of the Schmidt coefficients
(a non-decreasing monotone), $D^{(\left\{k\right\})}_n$ are based on their geometric mean.
Furthermore, $D^{(\left\{k\right\})}_n$ are able to reproduce the fundamental algebraic
invariants, at least for the small numbers of qubits ($N \le 4$) for which these
quantities are known.
It is hoped that $D^{(\left\{k\right\})}_n$ will be useful in the determination of AIs and the
classification of systems with $N \ge 5$ qubits.

From the form of $D^{(\left\{k\right\})}_n$ given in Eq. (\ref{dsd}), 
it is clear how to extend
the definition to parties with Hilbert spaces of greater dimension,
as the geometric mean construction is not dependent on this dimension
being two.
Finally we note that it is a simple step to introduce the average or minimum
of the quantities $D^{(\left\{k\right\})}_n$ and obtain a single EM for 
a given $n$ in the same way
has been done for the linear entropies \cite{sco04,ce04}.  The behaviour 
of such quantities is left for future work.

\section*{Acknowledgements}
I am grateful to C. W. J. Beenakker for discussions and advice.
This work was supported by the Dutch Science Foundation NWO/FOM.

\appendix
\section{Transformation properties of $D^{(\left\{k\right\})}_n$}

In this appendix we show that $D^{(\left\{k\right\})}_n$ is invariant under local unitary transformations.
We also consider the action of a POVM 
(positive operator--valued measurement) on the system.

\subsection{Invariance under unitary transformations of reduced qubits}
We first consider the invariance of $ D^{(\left\{k\right\})}_n$ with respect to
unitary transformations of a qubit belonging to $k_1,\ldots ,k_n$, which we take
to be the first without lack of generality.
We write the wave function as
\beq
  \ket{\Psi} =
  \sum_{X=0}^{l/2-1}
  \rb{
    \ket{0}\ket{\alpha_X}\ket{\mathbf{V}_{0,X}}
    +
    \ket{1}\ket{\alpha_X}\ket{\mathbf{V}_{1,X}}
  }
\eeq
where $\ket{\alpha_X}$ are basis states of the other qubits in  $\left\{k\right\}$,
and $\ket{\mathbf{V}_{(i=0,1),X}}$ are the same vectors as before, except that
we treat the index ($i$)
belonging to the first qubit separately from the rest ($X$).  We thus write
the set of vectors $\left\{\mathbf{V}\right\}$ as
$\left\{\mathbf{V}_{0,X},\mathbf{V}_{1,X}\right\}$, with
the vectors $\mathbf{V}_{0,X}$
to the left of $\mathbf{V}_{1,X}$.
We act on the first qubit with a general unitary operator $U$, giving the wave function
\beq
  U\ket{\Psi} =
  \sum_{X=0}^{l/2-1}
    \ket{0}\ket{\alpha_X}
    (U_{00} \ket{\mathbf{V}_{0,X}} + U_{10} \ket{\mathbf{V}_{1,X}})
  \nonumber\\
    +
    \ket{1}\ket{\alpha_X}
    (U_{01} \ket{\mathbf{V}_{0,X}} + U_{11} \ket{\mathbf{V}_{1,X}})
  \label{Upsi}
  .
\eeq

Define
\beq
  F_{\left\{j\right\}}(\ket{\Psi}) \equiv
  \bigwedge_{i=0}^{l-1} dx_{j_i}
  \rb{\left\{\mathbf{V}_{0,X},\mathbf{V}_{1,X}\right\}}
\eeq
in terms of which
\beq
  D^{(\left\{k\right\})}_n
  =
  l^2
  \left\{
    \sum_{j_0<\ldots <j_{l-1}}
     F_{\left\{j\right\}}(\ket{\Psi})   F_{\left\{j\right\}}(\ket{\Psi^*})
  \right\}^{2/l}
  .
\eeq
From Eq. (\ref{Upsi}), the
wedge product for the transformed wave function is
\beq
  F_{\left\{j\right\}}(U\ket{\Psi})
  =\bigwedge_{i=0}^{l-1} dx_{j_i}
  \rb{
  \left\{
    U_{00} \mathbf{V}_{0,X}+ U_{10} \mathbf{V}_{1,X}
  \right\},
  ~~~~
  \right.
  \nonumber\\
  \left.
  \left\{
    U_{01} \mathbf{V}_{0,X}+ U_{11} \mathbf{V}_{1,X}
  \right\}
  }
  .
\eeq
Since $\bigwedge dx$ is linear, and zero when any two of its arguments are
the same, we can write
\beq
  F_{\left\{j\right\}}(U\ket{\Psi})
  =\sum_{k_0,\ldots,k_{l-1}} \bigwedge_{i=0}^{l-1} dx_{j_i}
  \rb{
    \left\{
      U_{k_X 0} \mathbf{V}_{k_X,X}
    \right\},
  ~
  \right.
  \nonumber\\
  \left.
    \left\{
    U_{\bar{k}_X 1} \mathbf{V}_{\bar{k}_X,X}
    \right\}
  }
\eeq
where $k_X=0,1$ and $\bar{k}_X = (1+k_X)~\mathrm{mod}~2$.  We proceed
by rearranging the vectors in
the above expression such that all $\mathbf{V}_{0,X}$ stand to the left 
of $\mathbf{V}_{1,X}$, and collecting the
appropriate elements of $U$ with signs given by the antisymmetry of the wedge product.
The term that requires no interchange of vectors
acquires a forefactor $U_{00}^{l/2} U_{01}^{l/2}$, and there is only one such term.
There are $l/2$ terms that require a single
exchange of vectors. These terms have a forefactor
$U_{00}^{l/2-1}U_{10}U_{01}^{l/2-1}U_{11}$,
and acquire a minus sign due to the antisymmetry.  Proceeding similarly for all
the terms we arrive at
\beq
  F_{\left\{j\right\}}(U\ket{\Psi})
  &=&
    \sum_{k=0}^{l/2} {l/2 \choose k} (U_{00} U_{11})^{l/2-k} (-U_{01} U_{10})^k
  \nonumber \\
  && ~~~~~~~~~
  \times  \bigwedge_{i=0}^{l-1} dx_{j_i}
  \rb{\left\{\mathbf{V}_{0,X}\right\},\left\{\mathbf{V}_{1,X}\right\}}
  \nonumber\\
  &=&
  \rb{\mathrm{Det}U}^{l/2} F_{\left\{j\right\}}(\ket{\Psi})
\eeq
Therefore,
the effect of a unitary transformation on any of the reduced qubits is to multiply
each of the terms in $D^{(\left\{k\right\})}_n$ by $|\mathrm{Det}U|^2=1$.  Thus $D^{(\left\{k\right\})}_n$ is invariant under such
transformations.

This invariance also holds when we consider general transformations of
the entire $n$ qubit subsystem defined by the locus of $D^{(\left\{k\right\})}_n$.
The operation
of the most general $l\times l$ unitary operator on this $n$ qubit Hilbert
space multiplies
each term in $D^{(\left\{k\right\})}_n$ by $|\mathrm{Det}U|^{4/l}=1$, demonstrating the invariance as above.
Such invariance is not a requirement for being an EM, but it will be of use in the following.

\subsection{Invariance under unitary transformation of wedge-product}

The unitary invariance of the system under transformations of the
entire reduced qubit Hilbert space enables us to write the state
vector in a Schmidt decomposition
$\ket{\Psi}=\sum_{i=0}^{l-1} \ket{\phi_i}\ket{\mathbf{\tilde{V}}_i}$ 
without altering $D^{(\left\{k\right\})}_n$.
The vectors
$\ket{\mathbf{\tilde{V}}_i}$ are orthogonal but not normalised.
To demonstrate that $D^{(\left\{k\right\})}_n$ is invariant under local unitary
transformations of the qubits inside the wedge product,
we begin by writing
$D^{(\left\{k\right\})}_n$ as
\beq
 D^{(\left\{k\right\})}_n
 &=& l^2
 \left\{
   \sum_{j_0<\ldots <j_{l-1}}
   \bigwedge_{i=0}^{l-1} dx_{j_i}
   \rb{\left\{\mathbf{\tilde{V}}\right\}}
 \right.
  \nonumber\\
 &&~~~~~~~~~~~~~~~~~~~
 \left.
   \times
   \bigwedge_{i=0}^{l-1} dx_{j_i}
   \rb{\left\{\mathbf{\tilde{V}}^*\right\}}
  \right\}^{2/l}
  .
\eeq
We change the sum to include all values $\left\{j_i\right\}$, and write the
wedge-products
as tensors
\beq
 D^{(\left\{k\right\})}_n
 &=&  \frac{l^2}{l!}
 \left\{
   \sum_{j_0,\ldots,j_{l-1}}
   \rb{
     \bigwedge_{i=0}^{l-1} dx
     \rb{\left\{\mathbf{\tilde{V}}\right\}}
   }_{\left\{j_i\right\}}
  \right.
  \nonumber\\
  && ~~~~~~~~~~~~~~
  \left.
  \times
   \rb{
     \bigwedge_{i=0}^{l-1} dx
     \rb{\left\{\mathbf{\tilde{V}}^*\right\}}
   }_{\left\{j_i\right\}}
 \right\}^{2/l}
  .
\eeq
The tensor $\bigwedge_{i=0}^{l-1} dx \rb{\left\{\mathbf{\tilde{V}}\right\}}$ corresponds
to a sum of ordered $l$-tuples of vectors
\beq
  \bigwedge_{i=0}^{l-1} dx \rb{\left\{\mathbf{\tilde{V}}\right\}}
  = \sum_{\left\{k\right\}}\epsilon_{\left\{k\right\}}
  \left[
     \mathbf{\tilde{V}}_{k_0},\ldots,\mathbf{\tilde{V}}_{k_{l-1}}
  \right]
\eeq
with antisymmetric coefficients $\epsilon_{\left\{k\right\}}=\pm 1$ from the
determinant of Eq. (\ref{WPdef}).  Using this form, we have
\beq
  D^{(\left\{k\right\})}_n
  =  \frac{l^2}{l!}
  \left\{
    \sum_{j_0,\ldots,j_{l-1}}
    \rb{
      \sum_{\left\{k\right\}}\epsilon_{\left\{k\right\}}
      \left[
        \mathbf{\tilde{V}}_{k_0},\ldots,\mathbf{\tilde{V}}_{k_{l-1}}
      \right]
    }_{\left\{j_i\right\}}
  \right.
  \nonumber\\
  \left.
  \times
        \rb{
      \sum_{\left\{k'\right\}}\epsilon_{\left\{k'\right\}}
      \left[
        \mathbf{\tilde{V}}^*_{k'_0},\ldots,\mathbf{\tilde{V}}^*_{k'_{l-1}}
      \right]
    }_{\left\{j_i\right\}}
  \right\}^{2/l}
  .
\eeq
Summing over the $j$ indices, and recognising the scalar product of two vectors, we have
\beq
  D^{(\left\{k\right\})}_n
  =
  \frac{l^2}{l!}
  \left\{
    \sum_{\left\{k\right\}, \left\{k'\right\}}
    \epsilon_{\left\{k\right\}}\epsilon_{\left\{k'\right\}}
  \right.
  ~~~~~~~~~~~~~~~~~~~~~~~~~~~~
  \nonumber\\
  \left.
  \times
    \ew{\mathbf{\tilde{V}}_{k_0}|\mathbf{\tilde{V}}_{k'_0}}
    \ew{\mathbf{\tilde{V}}_{k_1}|\mathbf{\tilde{V}}_{k'_1}}
    \ldots
    \ew{\mathbf{\tilde{V}}_{k_{l-1}}|\mathbf{\tilde{V}}_{k'_{l-1}}}
  \right\}^{2/l}
  .
\eeq
We now use the orthogonality of $\ket{\mathbf{\tilde{V}}}$ from the Schmidt decomposition, and
that fact that
$\epsilon_{\left\{k\right\}}^2=1$ to write
\beq
  D^{(\left\{k\right\})}_n
  =
  l^2
  \left\{
    \prod_{i=0}^{l-1}
    \ew{\mathbf{\tilde{V}}_i|\mathbf{\tilde{V}}_i}
  \right\}^{2/l}
  .
\eeq
Thus we see that $D^{(\left\{k\right\})}_n$ is the product of Schmidt coefficients, and thus invariant
with respect to unitary transformations of the qubits inside the wedge-product.

\subsection{Action of POVM}

That $D^{(\left\{k\right\})}_n$ is an EM follows from the above results and the
argument given in the main text. Here we give
an alternative demonstration, based on that used by D\"{u}r {\it et al.} in establishing
the tangle as an EM \cite{dur00}, which provides justification for the power $2/l$
chosen in the definition of $D^{(\left\{k\right\})}_n$.

Any local protocol can be decomposed into POVMs 
acting on a single qubit.
As any POVM can be further decomposed into a sequence of two-outcome POVMs, we need only
to demonstrate the non-increasing of $D^{(\left\{k\right\})}_n$ under the action of a two-outcome POVM
to show that it is an EM.

Let the two elements of the POVM be $A_1$ and $A_2$ such
that $A_1^\dag A_1 + A_2^\dag A_2 =1$.
Using singular-value decompositions for these matrices, we have
$A_i = U_i X_i V$, where $U_i$, $V$
are unitary, $V$ is the same for both elements, and $X_{1,2}$
are the diagonal matrices $(a,b)$ and $(\sqrt{1-a^2}, \sqrt{1-b^2})$.

Consider the initial state $\ket{\psi}$, which possesses the measure $D^{(\left\{k\right\})}_n(\psi)$.
We write the (unnormalised) states obtained
by the action of the POVM on the state as $\ket{\widetilde{\phi}_i} = A_i \ket{\psi}$.
Normalising them,
we have $\ket{\phi_i} = \ket{\widetilde{\phi}_i}/\sqrt{p_i}$, where
$p_i \equiv \ew{\widetilde{\phi}_i|\widetilde{\phi}_i}$ with $p_1 + p_2=1$.

In analogy with the tangle, we wish to show that $(D^{(\left\{k\right\})}_n)^\nu$, $0 < \nu \le 1$ is
non-increasing,
on average, under the action of the POVM, i.e.
\beq
  \ew{(D^{(\left\{k\right\})}_n)^\nu} \le (D^{(\left\{k\right\})}_n)^\nu(\psi)
\eeq
with
\beq
 \ew{(D_n^{(\left\{k\right\})})^\nu} = p_1 (D^{(\left\{k\right\})}_n)^\nu(\phi_1)
  + p_2 (D^{(\left\{k\right\})}_n)^\nu(\phi_2)
\eeq
for all possible choices of POVM and states $\ket{\psi}$.
Since $D^{(\left\{k\right\})}_n$ is invariant under unitary transformations, we may omit the
 matrices $U_i$
from the decomposition of the POVM, such that
$D^{(\left\{k\right\})}_n(\phi_i) = D^{(\left\{k\right\})}_n(X_i V \psi / \sqrt{p_i})$.

Performing the POVM on one of the reduced qubits, we evaluate $\ew{(D^{(\left\{k\right\})}_n)^\nu}$
to be
\beq
  \ew{(D^{(\left\{k\right\})}_n)^\nu} =
  \left\{
    p_1 \rb{\frac{a^2 b^2}{p_1^2}}^\nu
    + p_2 \rb{\frac{(1-a^2)(1-b^2)}{p_2}}^\nu
  \right\}
  \nonumber\\
  ~~~~~~~~~~~~~~~
  \times
  (D_n^{(\left\{k\right\})})^\nu(\psi)
  \label{deta1}
  .
 \nonumber\\
\eeq
This is exactly the same dependence as found by D\"{u}r {\it et al.}
for the tangle \cite{dur00}. For all
$0 < \nu \le 1$ the forefactor in Eq. (\ref{deta1}) is not greater than one, and thus
$\ew{(D^{(\left\{k\right\})}_n)^\nu} \le (D^{(\left\{k\right\})}_n)^\nu(\psi)$, reiterating the conclusion that
$D^{(\left\{k\right\})}_n$ is an EM.
This result shows
that the choice of the power $2/l$ in the definition
of $D^{(\left\{k\right\})}_n$ is the natural choice, as it makes
$D^{(\left\{k\right\})}_n$ transform under local POVMs in the same way
as do the concurrence-squared and the tangle.



\end{document}